\documentclass[12pt]{article}
\usepackage[margin=1in]{geometry}
\usepackage{amsfonts,amsmath,amsthm,array,amssymb}
\usepackage[titletoc,title]{appendix}
\usepackage{hyperref}
\usepackage{natbib}
\usepackage{graphicx}
\usepackage{subfigure}
\usepackage{longtable}
\newtheorem{theorem}{Theorem}
\newtheorem{lemma}{Lemma}

\usepackage[margin=1in]{geometry}
\usepackage{setspace}
\usepackage{epstopdf}
\usepackage{float}

\usepackage{color}

\usepackage{epsfig}
\usepackage{figsize}
\usepackage{epsfig,longtable}

\date{}

\begin{document}

\title{Modeling the case of early detection of Ebola virus disease}

\author{
 Diego Chowell$^{1, 2,*}$, Muntaser Safan$^{1, 3, 4,*}$, and Carlos Castillo-Chavez$^{1}$\\
\footnotesize  $^{1}$Simon A Levin Mathematical, Computational and Modeling Sciences Center,\\
\footnotesize Arizona State University, Tempe, AZ, USA\\
\footnotesize  $^{2}$Center for Personalized Diagnostics, Biodesign Institute, Arizona State University, Tempe, AZ, USA\\
\footnotesize $^{3}$ Mathematics Department, Faculty of Science, Mansoura University, Mansoura, Egypt\\
\footnotesize $^{4}$Department of Mathematical Sciences, Faculty of Applied Sciences, Umm Al-Qura University, \\
\footnotesize Makkah Almukarramah, KSA\\
\footnotesize $^{*}$These authors contributed equally to this work
}
\maketitle

\begin{abstract}
The most recent Ebola outbreak in West Africa highlighted critical weaknesses in the medical infrastructure of the affected countries, including effective diagnostics tools, sufficient isolation wards, and enough medical personnel. Here, we develop and analyze a mathematical model to assess the impact of early diagnosis of pre-symptomatic individuals on the transmission dynamics of Ebola virus disease in West Africa in scenarios where Ebola may remain at low levels in the population. Our findings highlight the importance of implementing integrated control measures of early diagnosis and isolation. The mathematical analysis shows a threshold where early diagnosis of pre-symptomatic individuals, combined with a sufficient level of effective isolation, can lead to an epidemic control of Ebola virus disease. That is, the local erradication of the disease or the effective management of the disease at low levels of endemicity.

\end{abstract}

 \section{\protect\normalsize Introduction}

The Ebola viral strains are re-emerging zoonotic pathogens and members of the Filoviridae family consisting of five distinct species: \textit{Bundibugyo, Cotes d'Ivoire, Reston, Sudan}, and \textit{Zaire} with a high case-fatality rate in humans [\ref{lit:KugelmanJR}]. Filoviruses are long filamentous enveloped, non-segmented, single-stranded viruses, consisting of a negative-sense RNA genome [\ref{lit:BeechingNJ}]. Each Ebola species genome encodes seven linearly arranged genes: nucleoprotein (NP), polymerase cofactor (VP35), matrix protein (VP40), glycoprotein (GP), replication-transcription protein (VP30), matrix protein (VP24), and RNA-dependent RNA prolymerase (L)  [\ref{lit:BeechingNJ}]. While there are no proven effective vaccines or effective antiviral drugs for Ebola, containing an outbreak relies on contact tracing and on  early detection of infected individuals for isolation and care in treatment centers  [\ref{lit:BeechingNJ}]. The most recent Ebola outbreak in West Africa, which began in December 2013, due to the Zaire strain, demonstrated several weaknesses in the medical infrastructure of the affected countries, including the urgent need of effective diagnostics, which have a fundamental role in both disease control and case management.\\
 
The Ebola virus is transmitted as a result of direct contact with bodily fluids containing the virus  [\ref{lit:FauciAS}]. The virus enters via small skin lesions and mucus membranes where it is able to infect macrophages and other phagocytic innate immune cells leading to the production of a large number of viral particles  [\ref{lit:BeechingNJ}]. The macrophages, monocytes, and dendritic cells infected in the early stage of the disease serve to spread the virus throughout the organs, particularly in the spleen, liver, and lymph nodes  [\ref{lit:BeechingNJ}]. Consequently, critically ill patients display intensive viremia [\ref{lit:McElroyAK}]. Recognizing signs of Ebola viral disease is challenging because it causes common non-specific symptoms such as fever, weakness, diarrhea, and vomiting, and the incubation period typically lasts 5 to 7 days [\ref{lit:FauciAS}]. Therefore, functioning laboratories and effective point-of-care tests are critically needed in order to minimize transmission, allow better allocation of scarce healthcare resources, and increase the likelihood of success of antiviral treatments as they are developed [\ref{lit: FinkS}]. The puzzling case of Pauline Cafferkey, crtically ill, months after her recovery from Ebola, points to our lack of full understanding of Ebola infection and the likelihood of sustainable reservoirs, possible among the recovered.\\

There is an ongoing effort in place to improve Ebola diagnostics, primarily to detect the disease early. In our current state, the cost and difficulty of testing limit diagnostic facilities to small mobile laboratories or centralized facilities with turnaround times measured in days rather than in a few hours, meaning that diagnosis is largely used to confirm disease. Ebola diagnosis can be achieved in two different ways: measuring the host-specific immune response to infection (e.g. IgM and IgG antibodies) and detection of viral particles (e.g. ReEBOV Antigen Rapid Test Kit for VP40), or particle components in infected individuals (e.g. RT-PCR or PCR). The most general assay used for IgM and IgG antibody detection are direct ELISA assays. Table 1 illustrates recently manufactured kits and their corresponding assay name or type. Considering the physiological kinetics of the humoral immune system as well as impaired antigen-presenting cell function as a result of viral hemorrhagic fever, antibody titers are low in the early stages and often undetectable in severe patients prior to death [\ref{lit:IppolitoG}]. This leaves polymerase chain reaction (PCR) for antigen detection as a viable option for early diagnostic assays. PCR is a chemical reaction that amplifies pieces of a virus's genes floating in the blood by more than a millionfold, which makes detection of pre-symptomatic individuals likely identifiable. Indeed, a research article published in 2000, illustrates the power of this technology to detect Ebola virus in humans in the pre-symptomatic stage [\ref{lit:LeroyEM}]. In this study, 24 asymptomatic individuals who had been exposed to symptomatic Ebola patients were tested using PCR. Eleven of the exposed patients eventually developed the infection. Seven of the 11 tested positive for the PCR assay. And none of the other 13 did.

In this chapter, we extend the work presented in [\ref{lit: ChowellD}]. Here, we have developed and analyzed a mathematical model to evaluate the impact of early diagnosis of pre-symptomatic individuals on the transmission dynamics of Ebola virus disease in West Africa, under the assumption that the disease is maintained possibly at very low levels due to the deficiencies in health systems and our incomplete understanding of Ebola infection as illustrated by the case of  Pauline Cafferkey. Therefore, eliminating Ebola may require a more sustained and long-term effort that requires the use of models that include vital dynamics.

\begin{table}[H]
\footnotesize{
\begin{center}
\caption[Ebola molecular assays ]{Example of Ebola molecular assays and commercial lateral flow immunoassays.\\ PQ stards for prequalification.}
\begin{tabular}{lll}
\hline
\textbf{Company} & \textbf{Assay Name or Type} & \textbf{Regulatory Status} \\ \hline 
Altona Diagnostics (Germany) &  Ebolavirus RT-PCR Kit 1.0 & FDA Emergency Use \\
Roche (Switzerland) & LightMix Modular Ebola Virus Zaire & FDA EUA \\
CDC (USA) & CDC Ebola NP Real-time RT-PCR & FDA EUA \\
CDC (USA) & CDC Ebola VP40 Real-time RT-PCR &  FDA EUA  \\
US Department of Defense & DoD EZ1 Real-time RT-PCR & FDA EUA \\
BioMerieux (France) & BioFire Defense FilmArray Biothreat-E test &  FDA EUA \\
Cepheid (USA) & Xpert Ebola & FDA EUA \\
Atomic Energy Commission (France) & Ebola eZYSCREEN  & WHO PQ submitted \\
Chembio Diagnostics (USA) & Lateral flow immunoassay & WHO PQ submitted \\
Corgenix (USA) & ReEBOV Antigen Rapid Test & WHO PQ submitted \\
InTec (China) & Lateral flow immunoassay & WHO PQ submitted \\
Orasure (USA) & Lateral flow immunoassay & WHO PQ submitted \\ \hline
\end{tabular}
\label{tab:States}
\end{center}
}
\end{table}

 \section{\protect\normalsize Model formulation}
\begin{figure}[h]
\begin{centering}
\includegraphics[scale=0.65]{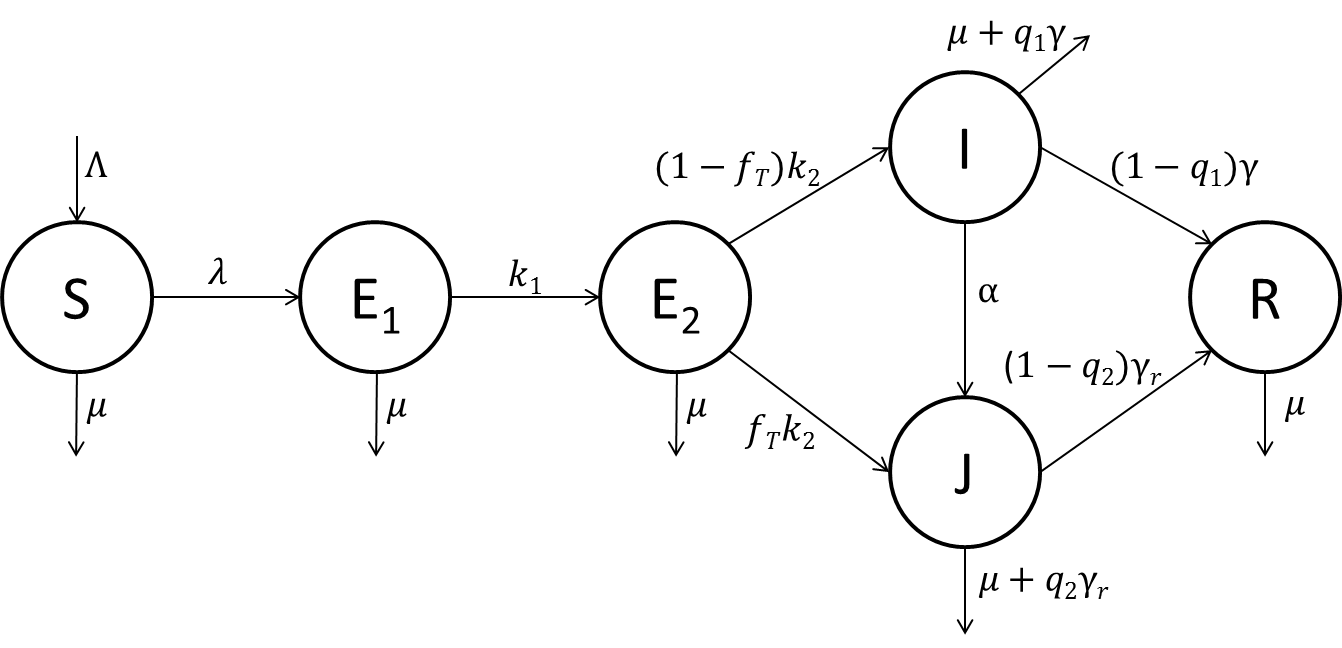}
\caption{Compartmental model showing the transition between model states. \label{Fig:Flowchart}}
\end{centering}
\end{figure}

The total population is assumed to be classified into six mutually independent subgroups: susceptible $S(t)$, non-detectable latent $E_1(t)$, detectable latent $E_2(t)$, infectious $I(t)$, isolated $J(t)$, and recovered $R(t)$ individuals. Table \ref{tab:States} shows the state variables and their physical meaning. The transition between all these states is shown in Figure \ref{Fig:Flowchart}. And model parameters and their description are presented in Table \ref{tab:ParamsDef}. Parameter values have been obtained from previous studies [\ref{lit: ChowellG1},\ref{lit: FasinaF}]. 

It is assumed that individuals are recruited (either through birth or migration)  into the susceptible class at a rate $\Lambda$ and die naturally with rate $\mu$. Susceptible individuals get infected due to successful contacts with infectious or not perfectly isolated infected individuals at rate $\lambda$. As a consequence, they become latent undetectable, who develop their state of infection to become latent detectable at rate $\kappa_1$. We assume that the latent detectable class represent individuals whose viral load is above the detection limit of the PCR-based diagnostic test [\ref{lit:LeroyEM},\ref{lit: QiuX}]. Latent detectable individuals either are diagnosed and get isolated with probability $f_T$ or develop symptoms to become infectious, who sequentially either get isolated at rate $\alpha$, or are removed from the system by recovery or Ebola-induced death at rate $\gamma$. It is assumed here that Ebola-induced deaths occur for the infectious individuals with probability $q_1$. Similarly, isolated individuals leave their class at rate $\gamma_r$, by either dying due to Ebola with probability $q_2$, or they get recovered and become immune. It is assumed that isolation is partially effective so that successful contacts with susceptible individuals may lead to infection with probability $r$; this parameter is a measure of isolation effectiveness of infectious individuals. Thus, the force of infection is given by
\begin{equation}
\lambda(t) = \frac{\beta[I(t) + (1-r)\ell J(t)]}{N(t) - r J(t)}.
\end{equation}


\begin{table}[H]
\footnotesize{
\begin{center}
\caption[Model State Variables]{Definition of model states.}
\vspace{2mm}
\begin{tabular}{ll}
\hline
\textbf{Variable} & \textbf{Description} \\ \hline 
$S(t)$&  Number of susceptible individuals at time $t$   \\
$E_1(t)$ & Number of latent undetectable individuals at time $t$  \\
$E_2(t)$ & Number of latent detectable individuals at time $t$ \\
$I(t)$ & Number of infectious individuals at time $t$\\
$J(t)$ & Number of isolated individuals at time $t$ \\
$R(t)$& Number of recovered individuals at time $t$ \\ \hline
\end{tabular}
\label{tab:States}
\end{center}
}
\end{table}

\begin{table}[H]
\footnotesize{
\begin{center}
\caption[Model Parameters]{Definition of Model Parameters.}
\vspace{2mm}
\begin{tabular}{llll}
\hline
\textbf{Parameter} & \textbf{Value} & \textbf{Unit} & \textbf{Description} \\ \hline 
$\Lambda$& $17182$ & $\frac{population}{day}$ & Recruitment rate  \\
$\beta$& $0.3335$ & ${day}^{-1}$ & Mean transmission rate   \\
$\mu$ & $4.98\times {10}^{-5}$ & ${day}^{-1}$& Natural death rate  \\
$\kappa_1$ & ${1}/{4}$ & ${day}^{-1}$ & Transition rate from undetectable to detectable latent state \\
$\kappa_2$ & ${1}/{3}$ & ${day}^{-1}$ & Exit rate of latent detectable individuals by either  \\
 & & & becoming infectious or moving to isolation state\\
$\gamma$ &${1}/{6}$ & ${day}^{-1}$ & Removal rate of infectious individuals by either recovery \\
 & & &  or Ebola-induced death  \\
$\gamma_r$& ${1}/{7}$ & ${day}^{-1}$  & Removal rate of isolated individuals by either recovery \\
 & & & or Ebola-induced death  \\ 
$\alpha$ & ${1}/{5}$ & ${day}^{-1}$ & Rate at which infectious individuals get isolated \\
$f_T$ & $0.25\in[0, 1]$ &  -- & Fraction of latent detectable individuals who are diagnosed \\
 & & & and get isolated \\
$q_1$ & 0.7 &  --  & Probability that an infectious individual dies due to Ebola\\
$q_2$ & 0.63 &  --  & Probability that an isolated individual dies due to Ebola \\
$r$ & $0.35\in[0, 1]$ &  --  & Effectiveness of isolation \\
$\ell$& $0.5\in[0, 1]$ &  --   & Relative transmissibility of isolated individuals\\ 
& & & with respect to infectious individuals \\ \hline
\end{tabular}
\label{tab:ParamsDef}
\end{center}
}
\end{table}

The assumptions mentioned above lead to the following model of equations
\begin{eqnarray}
\frac{dS}{dt} & = & \Lambda - \lambda S - \mu S,\nonumber\\
\frac{dE_1}{dt} & = & \lambda S - (\kappa_1 + \mu) E_1,\nonumber\\
\frac{dE_2}{dt} & = & \kappa_1 E_1 - (\kappa_2 + \mu) E_2,\nonumber\\
\frac{dI}{dt} & = & (1-f_T)\kappa_2 E_2 - (\alpha+\gamma + \mu) I, \label{model1}\\
\frac{dJ}{dt} & = & f_T \kappa_2 E_2 + \alpha I - (\gamma_r + \mu) J,\nonumber\\
\frac{dR}{dt} & = & (1-q_1) \gamma I + (1-q_2)\gamma_r J - \mu R \nonumber
\end{eqnarray}
where
\begin{equation*}
N(t) = S(t) + E_1(t) + E_2(t) + I(t) + J(t) + R(t)
\end{equation*}
is the total population size at time $t$. On adding all equations of system (\ref{model1}) together, we get
\begin{equation}
\frac{dN}{dt} = \Lambda - \mu N - q_1\gamma I - q_2 \gamma_r J.\label{Neq}
\end{equation}

%
%
%
%
%
%
%

\section{\protect\normalsize Model analysis}
 \subsection{\protect\normalsize Basic properties}
Since model (\ref{model1}) imitates the dynamics of human populations, all variables and parameters should be non-negative. Thus, following the approach shown in appendix A of [\ref{lit:HTh}], we show the following result.
\begin{theorem}
The variables of model (\ref{model1}) are non-negative for all time.
\end{theorem}

\begin{lemma}
The closed set 
\begin{equation*}
\Omega = \big\lbrace (S, E_1, E_2, I, J, R)\in \mathbb{R}_+^6: \frac{\Lambda}{\mu + q_1\gamma + q_2\gamma_r}\le {S + E_1 + E_2 + I + J + R} \le \frac{\Lambda}{\mu}  \big\rbrace
\end{equation*}
is positively invariant for model (\ref{model1}) and is absorbing.
\end{lemma}
\noindent Proof: Equation (\ref{Neq}) implies that
\begin{eqnarray}
\frac{dN}{dt} &\le& \Lambda - \mu N, \label{Neq1}\\
\frac{dN}{dt} &\ge& \Lambda - (\mu+q_1\gamma+q_2\gamma_r) N. \label{Neq2}
\end{eqnarray}
It follows from (\ref{Neq1}) that 
\begin{equation}
N(t) \le  \frac{\Lambda}{\mu} + \left(N(0) -\frac{\Lambda}{\mu} \right) e^{- \mu t}\label{ineq1}
\end{equation}
and from (\ref{Neq2}) that 
\begin{equation}
N(t)\ge\frac{\Lambda}{\mu + q_1\gamma + q_2\gamma_r} + \left(N(0) -\frac{\Lambda}{\mu + q_1\gamma + q_2\gamma_r} \right) e^{-(\mu + q_1\gamma + q_2\gamma_r)t}. \label{ineq2}
\end{equation}
If we assume $N(0) > \Lambda/\mu$, then $dN/dt < 0$ and therefore (based on inequality (\ref{ineq1})), $N(t)$ decreases steadily until reaching $\Lambda/\mu$ when $t$ tends to $\infty$. Similarly, if we assume $N(0) < \Lambda/(\mu + q_1\gamma + q_2\gamma_r)$, then $dN/dt > 0$ and therefore  (based on inequality (\ref{ineq2})), $N(t)$ increases steadily until reaching a maximum at $\Lambda/(\mu + q_1\gamma + q_2\gamma_r)$ when $t$ tends to $\infty$. It remains to check the case if $N(0)$ lies in the phase between $\Lambda/(\mu + q_1\gamma + q_2\gamma_r)$ and $\Lambda/\mu$. To this end, both inequalities (\ref{ineq1}) and (\ref{ineq2}) are combined together to get 
\begin{equation*}
\frac{\Lambda}{\mu + q_1\gamma + q_2\gamma_r} + \left(N(0) -\frac{\Lambda}{\mu + q_1\gamma + q_2\gamma_r} \right) e^{-(\mu + q_1\gamma + q_2\gamma_r)t}\le N(t) \le \frac{\Lambda}{\mu} + \left(N(0) -\frac{\Lambda}{\mu} \right) e^{- \mu t}.
\end{equation*}
On taking the limit when $t$ tends to $\infty$, we find that $N(t)$ remains within the same phase. Thus, the set $\Omega$ is positively invariant and absorbing.

 \subsection{\protect\normalsize Equilibrium analysis}
 \subsubsection*{\protect\normalsize Ebola-free equilibrium and the control reproduction number $\mathcal{R}_c$}
It is easy to check that model (\ref{model1}) has the Ebola-free equilibrium 
\begin{equation}
E_0 = \left(\frac{\Lambda}{\mu}, 0, 0, 0, 0, 0\right)^{\prime}
\end{equation}
where the prime `` ${}^{\prime}$ '' means vector transpose.\newline 
\indent The basic reproduction number,  $\mathcal{R}_0$,  is a measure of the average number of secondary cases produced by a typical infectious individual during the entire course of infection in a completely susceptible population and in the absence of control interventions [\ref{lit:  BrauerF},\ref{lit: AndersonRM}]. On the other hand, the control reproduction number, $\mathcal{R}_c$, quantifies the potential for infectious disease transmission in the context of a partially susceptible population due to the implementation of control interventions. When $\mathcal{R}_c > 1$, the infection may spread in the population, and the rate of spread is higher with increasingly high values of $\mathcal{R}_c$. If $\mathcal{R}_c < 1$, infection cannot be sustained and is unable to generate an epidemic.  For our model, $\mathcal{R}_c$  is computed using the next generation matrix approach shown in [\ref{lit:PVDDJW2002}].  Accordingly, we compute the matrices $\mathbf{F}$ (for the new infection terms) and $\mathbf{V}$ (for the
transition terms) as

\begin{eqnarray*}
  \mathbf{F} = \left(\begin{array}{cccc}
    0 & 0 & \beta & (1-r) \ell \beta \\
    0 & 0 & 0 & 0 \\
    0 & 0 & 0 & 0 \\
    0 & 0 & 0 & 0 \\
  \end{array}\right) \quad, \quad
   \mathbf{V} = \left(\begin{array}{cccc}
    \kappa_1 + \mu & 0 & 0 & 0 \\
    -\kappa_1 & \kappa_2 + \mu  & 0 & 0\\
    0 & -(1-f_T) \kappa_2 & \alpha+\gamma+\mu & 0\\
    0 & - f_T \kappa_2  & - \alpha & \gamma_r + \mu\\
  \end{array}\right).
\end{eqnarray*}
\noindent Thus, the control reproduction number is given by
\begin{eqnarray}
\mathcal{R}_c & = &\rho(\mathbf{F}\mathbf{V}^{-1}) = \frac{\kappa_1 \kappa_2 \beta [(1-f_T) (\mu + \gamma_r) + (1-r)\ell(\alpha+f_T(\gamma+\mu))]}{(\kappa_1 + \mu)(\kappa_2 + \mu)(\alpha+\gamma+\mu)(\gamma_r + \mu)} \nonumber\\
& = & \frac{\kappa_1\kappa_2\beta}{(\kappa_1 + \mu)(\kappa_2 + \mu)(\alpha+\gamma+\mu)}\left[1 - f_T + (1-r) \ell \left( \frac{\alpha}{\gamma_r+\mu} + f_T  \frac{\gamma + \mu}{\gamma_r+\mu} \right) \right]\nonumber\\
& = & \mathcal{R}_0\left[1-\frac{\alpha}{(\alpha+\gamma+\mu)}\right] \left[1 - f_T + (1-r) \ell \left( \frac{\alpha}{\gamma_r+\mu} + f_T  \frac{\gamma + \mu}{\gamma_r+\mu} \right) \right]\label{R0eq}
\end{eqnarray}
where $\rho$ is the spectral radius (dominant eigenvalue in
magnitude) of the matrix $\mathbf{F}\mathbf{V}^{-1}$ and 
\begin{equation}
\mathcal{R}_0 = \frac{\kappa_1\kappa_2\beta}{(\kappa_1 + \mu)(\kappa_2 + \mu)(\gamma+\mu)}
\end{equation}
is the basic reproduction number for the model.

\indent The local stability of the Ebola-free equilibrium, $E_0$, for values of $\mathcal{R}_c < 1$ is established based on a direct use of Theorem 2 in [\ref{lit:PVDDJW2002}]. We summarize our result in the following lemma.
\begin{lemma}
The Ebola-free equilibrium $E_0$ of model (\ref{model1}) is locally asymptotically stable if and only if $\mathcal{R}_c < 1$.
\end{lemma}

\subsubsection*{\protect\normalsize Ebola-endemic equilibrium}
On putting the derivatives in the left hand side of (\ref{model1}) equal zero and solving the resulting algebraic system with respect to the variables $\bar{S}, \bar{E}_1, \bar{E}_2, \bar{I}, \bar{J}$, and $\bar{R}$, we obtain
 \begin{eqnarray}
\bar{S} & = & \frac{\Lambda}{\bar\lambda + \mu},\nonumber\\
\bar{E}_1 & = & \frac{\Lambda}{\bar\lambda + \mu} \cdot \frac{\bar\lambda}{\kappa_1 + \mu},\nonumber\\
\bar{E}_2 & = & \frac{\kappa_1}{\kappa_2 + \mu}\cdot \frac{\Lambda}{\bar\lambda + \mu} \cdot \frac{\bar\lambda}{\kappa_1 + \mu},\nonumber \\
\bar{I} & = & \frac{(1-f_T)\kappa_2}{\alpha+\gamma + \mu} \cdot \frac{\kappa_1}{\kappa_2 + \mu}\cdot \frac{\Lambda}{\bar\lambda + \mu} \cdot \frac{\bar\lambda}{\kappa_1 + \mu},\label{eqvar} \\
\bar{J} & = &  \frac{\kappa_1}{\kappa_2 + \mu}\cdot \frac{\Lambda}{\bar\lambda + \mu} \cdot \frac{\bar\lambda}{\kappa_1 + \mu} \cdot \frac{\kappa_2}{\gamma_r + \mu} \left[f_T + (1-f_T) \frac{\alpha}{\alpha+\gamma + \mu} \right], \nonumber\\
\bar{R} & = & \frac{1}{\mu}[(1-q_1)\gamma I + (1-q_2) \gamma_r J]\nonumber
\end{eqnarray}
where
\begin{equation}
\bar\lambda = \frac{\beta(I + (1-r)\ell \bar{J})}{\bar{N} - r \bar{J}}\label{lambda}
\end{equation}
is the equilibrium force of infection. On substituting from (\ref{eqvar}) into (\ref{lambda}) and simplifying (with the assumption that $\lambda \ne 0$), we get
 \begin{equation}
\bar\lambda = \frac{\mu(\mathcal{R}_c - 1)}{1 - Term}
\end{equation}
where 
 \begin{equation*}
Term = \frac{\kappa_1 \kappa_2 [q_1(1-f_T)\gamma(\gamma_r + \mu) + (r\mu + q_2\gamma_r)(f_T(\gamma + \mu) + \alpha)]}{(\kappa_1 + \mu)(\kappa_2 + \mu)(\alpha+\gamma+\mu)(\gamma_r + \mu)}.
\end{equation*}
Hence, the Ebola-endemic equilibrium is unique and we show the following lemma.
 \begin{lemma}
 Model (\ref{model1}) has a unique endemic equilibrium that exists if and only if $\mathcal{R}_c > 1$.
 \end{lemma}
 
%
%
%
\subsection{\protect\normalsize Normalized sensitivity analysis on $\mathcal{R}_c$ }\label{sensitivity}

In considering the dynamics of the Ebola system (\ref{model1}), we conduct normalized sensitivity analysis on $\mathcal{R}_c$ to determine the impact of parameter perturbations on the transmission dynamics of the system. By computing the normalized sensitivity indices, we consider the percent change in the output with respect to a percent change in the parameter input. Those parameters with the largest magnitude of change impact the compartment model the most; the sign indicates whether the change produces an increase or a decrease on $\mathcal{R}_c$.\newline
\indent The normalized sensitivity indices for $\mathcal{R}_c$ are calculated by taking the partial derivative of $\mathcal{R}_c$ with respect to each parameter and multiply the derivative with the ratio of the parameter to $\mathcal{R}_c$. This value represents the percent change in $\mathcal{R}_c$ with respect to a 1\% change in the parameter value [\ref{lit: CaswellH}]. \newline
 
\vspace{-5mm}
\begin{table}[h!]
\caption{Percent change in $\mathcal{R}_c$ with respect to a $1\%$ change in the parameter value, for a low and a high isolation effectiveness $r$, and a low and a high value of $f_T$, while keeping the other parameter values as presented in Table \ref{tab:ParamsDef}.} \label{r0change1}
\vspace{-6mm}
\begin{center}
\begin{tabular}{|c|c|c|c|c|c|c|c|c|}
\hline
&Parameter & $\beta$ & $r$ & $\ell$ & $\gamma_r$ & $\gamma$ & $\alpha$ & $f_T$ \\
\hline

 &\% change & 1\% & -0.23\% & 0.423\%  & -0.423\% & -0.382\% & -0.195\% & -0.119\% \\
 $f_T = 0.25$& for $r = 0.35$ &  &  &    &   &   &   &  \\
\cline{2-9}
&\% change & 1\% & -1.014\% & 0.053\%  & -0.053\% & -0.445\% & -0.501\% & -0.306\% \\
  & for $r = 0.95$ &  &  &    &   &   &   &  \\
\hline
&\% change & 1\% & -0.402\% & 0.747\%  & -0.747\% & -0.167\% & -0.086\% & -0.471\% \\
 $f_T = 0.75$& for $r = 0.35$ &  &  &    &   &   &   &  \\
\cline{2-9}
&\% change & 1\% & -3.521\% & 0.185\%  & -0.185\% & -0.383\% & -0.431\% & -2.373\% \\
  & for $r = 0.95$ &  &  &    &   &   &   &  \\
\hline
\end{tabular}
\end{center}
\end{table}

\begin{figure}[h!]
\begin{center}
\includegraphics[scale=0.57]{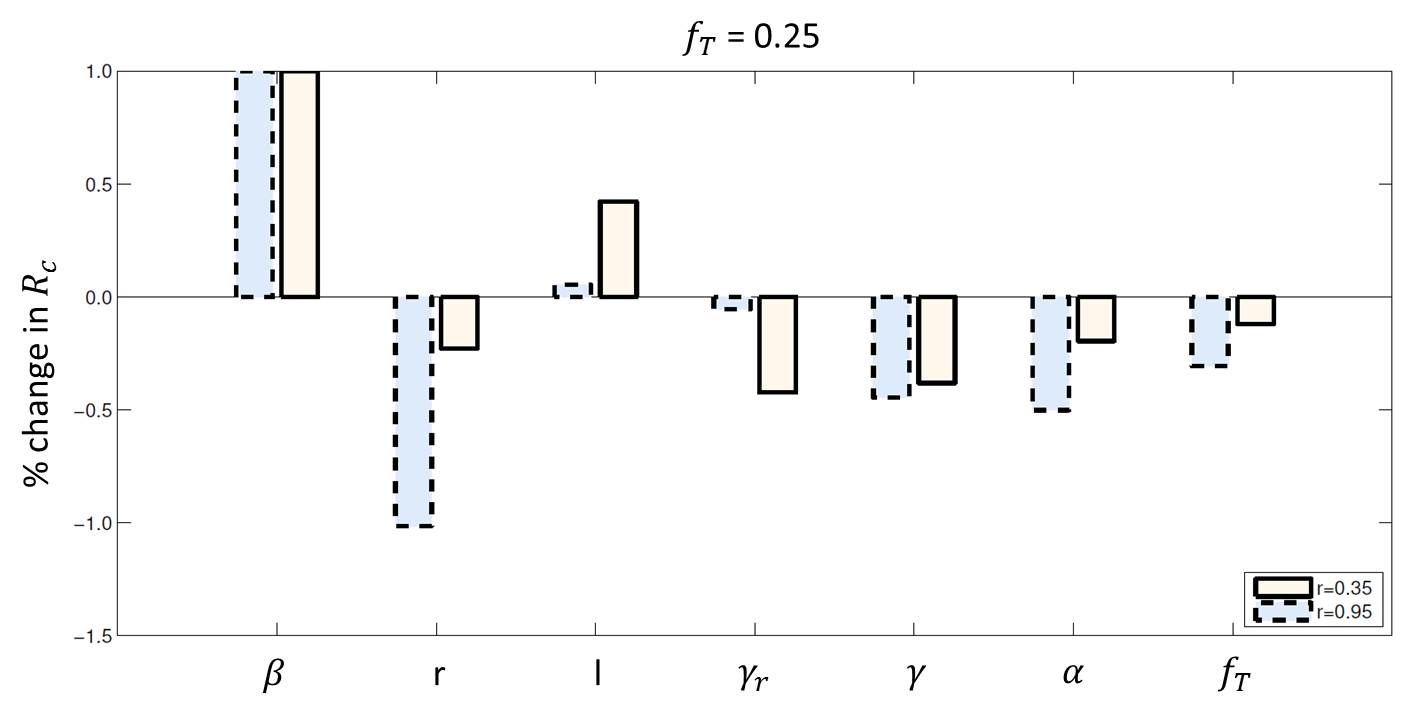}
\caption{Percent change in $\mathcal{R}_c$ with respect to a $1\%$ change in the parameter value, for a low value of $f_T$ ($f_T = 0.25$) and two different levels of isolation effectiveness ($r = 0.35$ and $r = 0.95$). The other parameter values are kept as shown in Table \ref{tab:ParamsDef}.} \label{graphofR01}
\end{center}
\end{figure}

%


\indent We use the parameters values from Table \ref{tab:ParamsDef} to study the sensitivity of $\mathcal{R}_c$ to each parameter. We compute normalized sensitivity analysis on all parameters, but we just consider the impact of parameters that are the most sensitive: $\beta, r, \ell, \gamma_r, \gamma, \alpha$, and $f_T$. The other parameters ($\mu, \kappa_1$, and  $\kappa_2$)  have a very low impact, namely less than $0.001\%$. The numerical simulations to the sensitivity of $\mathcal{R}_c$ with respect to each of the most sensitive parameters are given in Table \ref{r0change1}, for two different levels of isolation effectiveness ($r = 0.35$ and $r = 0.95$) and two values of $f_T$ ($f_T = 0.25$ and $f_T = 0.75$), which is the fraction of pre-symptomatic individuals diagnosed and isolated. The other parameter values are kept as shown in Table \ref{tab:ParamsDef}.
\newline
\indent A graphical illustration of the numerical results for the scenario when $f_T = 0.25$ and the two levels of isolation effectiveness ($r = 0.35$ and $r = 0.95$) is given in Figure \ref{graphofR01}. In the case of high isolation effectiveness ($r = 0.95$), simulations show that both the removal rate, $\gamma_r$, of isolated individuals and the relative transmissibility parameter $\ell$ of isolated individuals with respect to infectious individuals are the least sensitive parameters (with $0.053 \%$ change of $\mathcal{R}_c$), while the parameter of isolation effectiveness, $r$, is the most sensitive one, where a $1\%$ increase in $r$ causes a $1.014 \%$ reduction in the value of $\mathcal{R}_c$. Also, the rate at which infectious individuals get isolated, $\alpha$, and the fraction of pre-symptomatic individuals detected and isolated, $f_T$, impact negatively on the level of $\mathcal{R}_c$, where a $1 \%$ percent increase in the value of $f_T$ causes approximately a $0.31\%$ decline in the value of the reproduction number $\mathcal{R}_c$. Thus, as pre-symptomatic individuals are diagnosed and as isolation is highly effective, the number of available infectious individuals who are capable of transmitting Ebola decreases and therefore, the reproduction number decreases. Also, the removal (by recovery or Ebola-induced death) rate $\gamma$ of infectious individuals affects negatively on $\mathcal{R}_c$. Hence, for the case of highly effective isolation, the parameters concerning early diagnosis and isolation have a significant impact on the reproduction number. 

%

\indent This percent impact of the parameters on $\mathcal{R}_c$ remains so as long as isolation is highly effective. However, if the effectiveness of isolation is low, in the sense that all parameter values are kept the same except the value of the parameter $r$, which is reduced to $0.35$, then we get the results presented in Table \ref{r0change1} and Figure \ref{graphofR01}. In this case, both the relative transmissibility $\ell$ and the removal rate of isolated individuals, $\gamma_r$, are the second most sensitive parameters, after $\beta$ which is the most impactful one. Also, $\ell$ became more sensitive than $r$. The implication is that, when isolation is less effective, there exists the possibility for isolated people to make successful contacts with susceptible individuals and therefore the possibility of causing new infections increases. This causes an increase in the reproduction number. Also, it is noted that the effect of $f_T$ and $\alpha$ is reduced, which means that diagnosing and isolating infected individuals becomes a weak strategy if the effectiveness of isolation is low.
 
 %
\begin{figure}[h!]
\begin{center}
\includegraphics[scale=0.585]{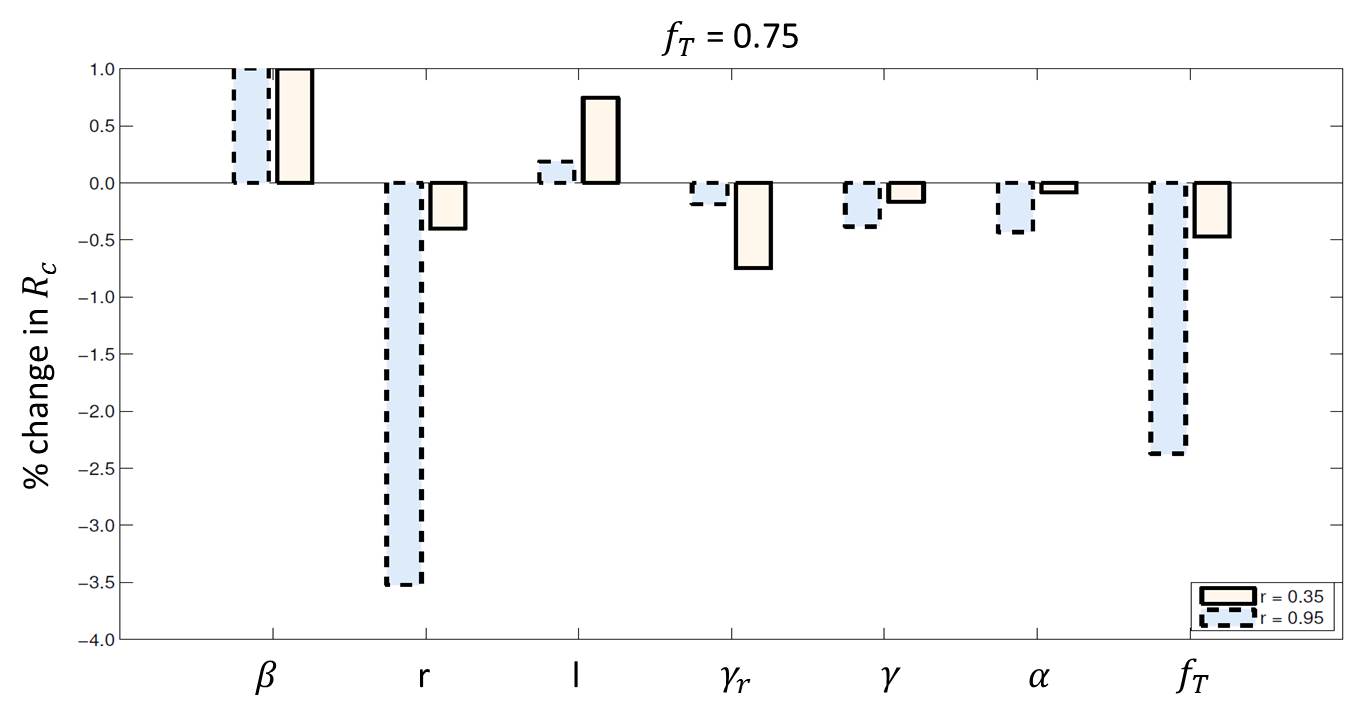}
\caption{Percent change in $\mathcal{R}_c$ with respect to a $1 \%$ change in the parameter value, for a high value of $f_T$ ($f_T = 0.75$) and two different levels of isolation effectiveness ($r = 0.35$ and $r = 0.95$). The other parameter values are kept as shown in Table \ref{tab:ParamsDef}.} \label{graphofR02}
\end{center}
\end{figure}

On repeating the previous analyses, but this time for a higher value of $f_T$ ($f_T = 0.75$), we obtain the results shown in Table  \ref{r0change1}, which are also illustrated in Figure \ref{graphofR02}. In comparison to the scenario when $f_T = 0.25$, the simulations show that increasing the fraction of pre-symptomatic individuals who are diagnosed and isolated,  $f_T$, increases the percent impact of the parameters $r, \ell, \gamma_r,$ and $f_T$, and decreases the percent impact of the parameters $\gamma$ and $\alpha$, on the value of the control reproduction number $\mathcal{R}_c$.


\subsection{\protect\normalsize Impact of early detection and isolation on the value of $\mathcal{R}_c$}

 \begin{figure}[H]
\centering
\includegraphics[scale=0.55]{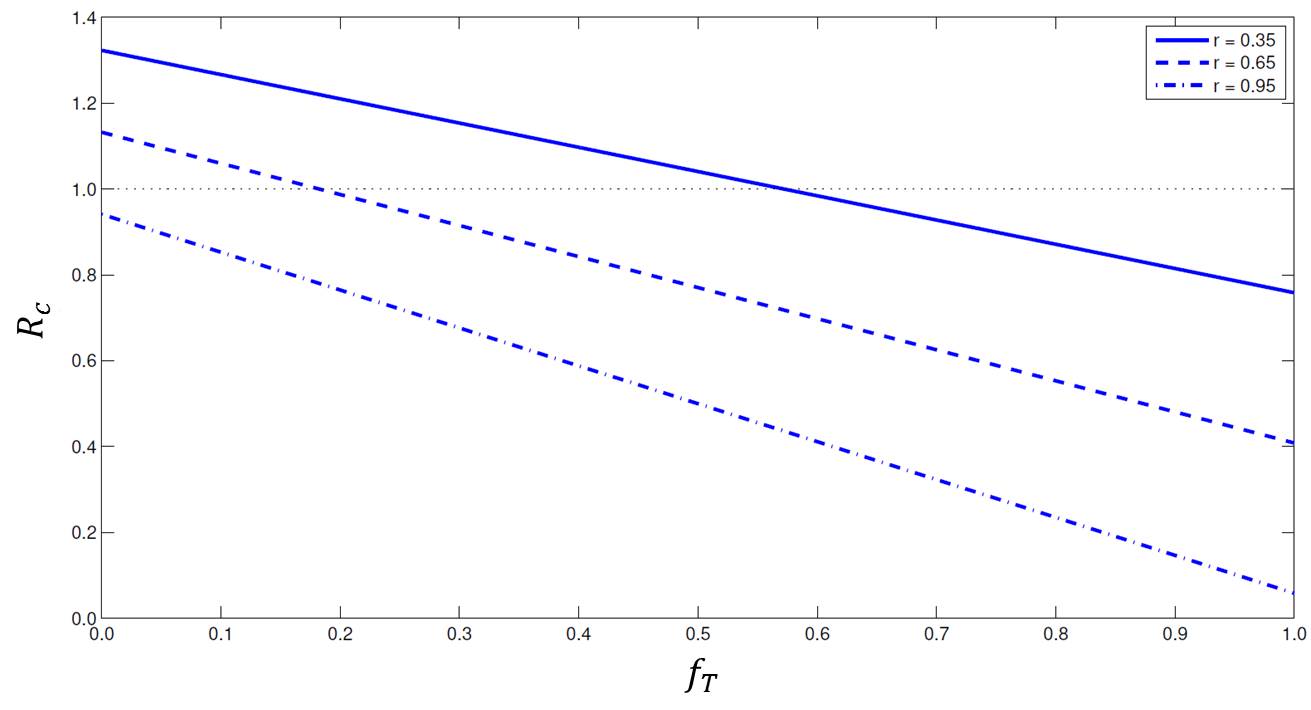}
\caption{Impact of early detection of pre-symptomatic individuals on the value of $\mathcal{R}_c$.}
\label{fig:R0fT}
\end{figure}

To study the impact of early detection of pre-symptomatic individuals and isolation on the reproduction number, we first depict $\mathcal{R}_c$ as a function of $f_T$, for different levels of isolation effectiveness $r$. Figure \ref{fig:R0fT} shows that the control reproduction number declines as the proportion, $f_T$, of pre-symptomatic individuals, who get diagnosed and isolated, increases.  Simulations are done using parameter values from Table \ref{tab:ParamsDef}, but for three different values of $r$. It further shows that the curve corresponding to a low and an intermidate value of isolation effectivenes $r$ (e.g. $r = 0.35$ for the solid curve and $r = 0.65$ for the dashed curve) hits $\mathcal{R}_c = 1$ at some critical value of $f_T$ (say $f_T^{\star}$), while for the high value of $r$ ($r = 0.95$), it never hits the critical threshold $\mathcal{R}_c = 1$, as the curve is totally below the critical threshold. This indicates that for a high effectiveness of isolation, the control reproduction number is less than one and therefore the infection dies out. Analytically, the exact form of $f_T^{\star}$ is 
\begin{equation}
f_T^{\star} = \left[ 1 + (1-r)\ell \frac{\alpha}{\gamma_r + \mu} - \frac{1}{\mathcal{R}_0} \left(1 + \frac{\alpha}{\gamma+\mu}\right) \right] / \left[ 1 - \frac{(1-r)\ell(\gamma + \mu)}{\gamma_r + \mu} \right].  \label{fTcond1}
\end{equation}
The critical proportion $f_T^{\star}$ represents the minimum proportion of pre-symptomatic individuals who are detected and get isolated to ensure an effective control of Ebola. This critical value remains feasible as long as the following inequality holds
\begin{equation}
(1-r)\ell < \frac{\gamma_r + \mu}{(\gamma + \mu)\mathcal{R}_0}. \label{fTcond20}
\end{equation}
If we keep all parameters fixed except $r$, then condition (\ref{fTcond20}) could be rewritten in a more convenient form
\begin{equation}
r > 1- \frac{\gamma_r + \mu}{\ell (\gamma + \mu) \mathcal{R}_0}. \label{fTcond3}
\end{equation}
This gives the minimum level of effectiveness of isolation required to obtain an isolation and early diagnosis-based control strategy for Ebola tranmission. 

 \begin{figure}[H]
\centering
\includegraphics[scale=0.55]{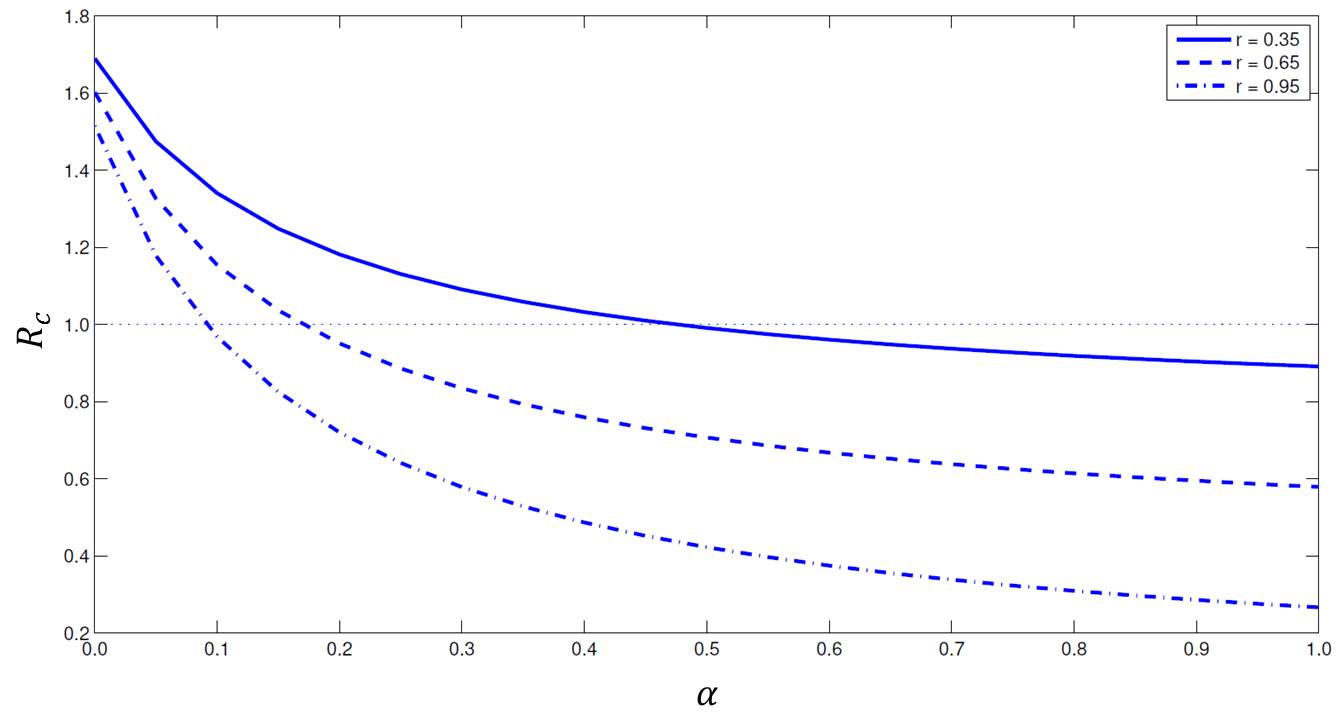}
\caption{Impact of isolating infectious individuals on the value of $\mathcal{R}_c$.}
\label{fig:R0alpha}
\end{figure}
Now, we could also ask a similar question on the role of isolating infectious individuals to contain Ebola transmission. Figure \ref{fig:R0alpha} shows the impact of changing the rate at which infectious individuals get isolated, $\alpha$, on $\mathcal{R}_c$, for the same three different levels of isolation effectivenes, as used above. The analysis shows that it is possible to control the epidemic if and only if $\alpha > \alpha^\star$, where
\begin{equation}
\alpha^\star = \frac{[ (1-f_T)(\gamma_r + \mu)(\gamma+\mu) + (1-r)\ell f_T (\gamma+\mu)^2]\mathcal{R}_0 -  (\gamma_r + \mu)(\gamma+\mu) }{(\gamma_r + \mu) - \ell (1-r) \mathcal{R}_0(\gamma + \mu)}
\end{equation}
%
and with the implementation of condition (\ref{fTcond20}).

%
%
%
%
%
%
%
%
%
%

\section{\protect\normalsize Discussion and conclusion}
\label{discussion}
The Ebola epidemic has shown us major weaknesses not only in health systems in West Africa, but also in our global capacity to respond early to an outbreak with effective diagnositc capacities. After multiple outbreaks of infectious diseases, from severe acute respiratory syndrome (SARS) to Middle East respiratory syndrome coronavirus (MERS-CoV), we still do not have effective diagnostic tools to rapidly respond to a number of potential epidemics. The main reason why we lack of such diagnostic preparedness against infectious diseases is because of the lack of a financed global strategy that can be implemented ahead, rather than during an epidemic. This strategy must primarily focus on two critical aspects: First, a continuous intereaction between the field to detect small outbreaks and collect samples, and reference laboratories with advanced sequencing tools to identify the pathogen. Second, the need of assay development for early diagnosis, their regulatory approval, and a plan of implementation in anticipation of an outbreak.\\ 
Here, motivated by some studies showing that PCR assay can detect Ebola virus in both humans and non-human primates during the pre-symptomatic stage [\ref{lit:LeroyEM},\ref{lit: QiuX}], we have developed and analyzed a mathematical model calibrated to the transmission dynamics of Ebola virus disease in West Africa to evaluate the impact of early diagnosis of pre-symptomatic infections. In the absence of effective treatments and vaccines, our results show the importance of implementing integrated control measures of early diagnosis and isolation. Importantly, our analysis identifies a threshold where early diagnosis of pre-symptomatic individuals, combined with a sufficient level of effective isolation, can lead to an epidemic control of Ebola virus disease. Furthermore, the need to incorporate vital dynamics is justified by our still limited understanding of Ebola infection including whether or not Ebola virus may persist among recovered individuals. The use of $\mathcal{R}_c$ in this context reflect our view that control measures should be sustainable and not just in response to an outbreak.


\section*{\protect\normalsize Acknowledgments}
We thank Benjamin Katchman for the helpful discussions about the different types of molecular diagnostics of Ebola.

\section*{\protect\normalsize References} 

\begin{enumerate}

\item{\normalsize\label{lit:KugelmanJR} Kugelman J. R., Sanchez-Lockhart M., Andersen K. G., Gire S., Park D. J., Sealfon R., Lin A. E., Wohl S., Sabeti P. C., Kuhn J. H., Palacios G. F. 2015. Evaluation of the potential impact of Ebola virus genomic drift on the efficacy of sequence-based candidate therapeutics. mBio 6(1):e02227-14. doi:10.1128/mBio.02227-14.}

\item{\normalsize\label{lit:BeechingNJ} Beeching, N. J., Fenech, M., Houlihan, C. F. (2014). Ebola virus disease. BMJ, 349, g7348.}

\item{\normalsize\label{lit:FauciAS} Fauci, A. S. (2014). Ebola$-$underscoring the global disparities in health care resources. New England Journal of Medicine, 371(12), 1084-1086.}

\item{\normalsize\label{lit:McElroyAK} McElroy, A. K., Erickson, B. R., Flietstra, T. D., Rollin, P. E., Nichol, S. T., Towner, J. S., Spiropoulou, C. F. (2014). Ebola hemorrhagic fever: novel biomarker correlates of clinical outcome. Journal of Infectious Diseases, jiu088.}

\item{\normalsize\label{lit:KortepeterMG} Kortepeter, M. G., Bausch, D. G., Bray, M. (2011). Basic clinical and laboratory features of filoviral hemorrhagic fever. Journal of Infectious Diseases, 204(suppl 3), S810-S816.}

\item{\normalsize\label{lit:IppolitoG} Ippolito, G., Feldmann, H., Lanini, S., Vairo, F., Di Caro, A., Capobianchi, M. R., Nicastri, E. (2012). Viral hemorrhagic fevers: advancing the level of treatment. BMC medicine, 10(1), 31.}

\item{\normalsize\label{lit:LeroyEM} Leroy, E. M., Baize, S., Volchkov, V. E., Fisher-Hoch, S. P., Georges-Courbot, M. C., Lansoud-Soukate, J., ...  McCormick, J. B. (2000). Human asymptomatic Ebola infection and strong inflammatory response. The Lancet, 355(9222), 2210-2215.}

\item{\normalsize\label{lit:HTh} Thieme, H. R.: Mathematics in Population Biology. Princeton University Press, 2003.}

\item{\normalsize\label{lit:PVDDJW2002} van den Driessche, P., Watmough, J.: Reproduction numbers and sub-threshold endemic equilibria for
compartmental models of disease transmission. {\it Math Biosci} {\bf 180} (2002), 29-48.}

\item{\normalsize\label{lit: CaswellH} Caswell, H. (2001). Matrix population models second edition. Sunderland, MA: Sinauer Associates.}


\item{\normalsize\label{lit: QiuX} Qiu, X, Wong, G, Audet, J et al. Reversion of advanced Ebola virus disease in nonhuman primates with ZMapp. Nature. 2014; 514: 47–53}

\item{\normalsize\label{lit: FinkS} Fink, S. Ebola drug aids some in a study in West Africa. The New York Times, Feb. 4, 2015.}

\item{\normalsize\label{lit: ChowellD} Chowell D, Castillo-Chavez C, Krishna S, Qiu X, and Anderson KS. Modelling the effect of early detection of Ebola. The Lancet Infectious Diseases. 2005. 15(2), 148-149.}

\item{\normalsize\label{lit: BrauerF} Brauer F, Castillo-Chavez C. Mathematical models in population biologyand epidemiology. Springer (2011).}

\item{\normalsize\label{lit: AndersonRM} Anderson RM, May RM. Infectious diseases of humans. Oxford: Oxford university press (1991).}

\item{\normalsize\label{lit: ChowellG1} Chowell G and Nishiura H. Transmission dynamics and control of Ebola virus disease (EVD): a review. BMC Med. 2014; 12: 196}

\item{\normalsize\label{lit: FasinaF} Fasina F, Shittu A, Lazarus D, et al. Transmission dynamics and control of Ebola virus disease outbreak in Nigeria, July to September 2014. Euro Surveill 2014; 19: 20920.}

\end{enumerate}

\end{document}